\documentclass{osa-article}

\journal{oe}
\articletype{Research Article}
\usepackage{lineno}
\usepackage{threeparttable}
\usepackage{multirow}
\usepackage{subcaption}
\begin{document}

\title{Ultrafast laser stress figuring for accurate deformation of thin mirrors}

\author{Brandon D. Chalifoux,\authormark{1,*} Kevin A. Laverty,\authormark{1} and Ian J. Arnold\authormark{1}}

\address{\authormark{1}James C. Wyant College of Optical Sciences, University of Arizona, 1630 E. University Blvd., Tucson AZ 85719, USA}

\email{\authormark{*}bchal@arizona.edu} %% email address is required



\begin{abstract}
Fabricating freeform mirrors relies on accurate optical figuring processes capable of arbitrarily modifying low-spatial frequency height without creating higher-spatial frequency errors. We present a scalable process to accurately figure thin mirrors using stress generated by a focused ultrafast laser. We applied ultrafast laser stress figuring (ULSF) to four thin fused silica mirrors to correct them to 10-20 nm RMS over 28 Zernike terms, in 2-3 iterations, without significantly affecting higher-frequency errors. We measured the mirrors over a month and found that dielectric-coated mirrors were stable but stability of aluminum-coated mirrors was inconclusive. The accuracy and throughput for ULSF is on par with existing deterministic figuring processes, yet ULSF doesn't significantly affect mid-spatial frequency errors, can be applied after mirror coating, and can scale to higher throughput using mature laser processing technologies. ULSF offers new potential to rapidly and accurately shape freeform mirrors.  
\end{abstract}

%%%%%%%%%%%%%%%%%%%%%%%%%%  body  %%%%%%%%%%%%%%%%%%%%%%%%%%
\section{Introduction}
Freeform mirrors have non-rotationally symmetric surface shapes that open new design spaces for optical systems with capabilities not possible with spherical or conic surfaces: for example, larger field of view, fewer components, and smaller mass and volume \cite{schiesser_effect_2019}. However, the lack of symmetry that makes freeform mirrors so powerful for design complicates fabrication \cite{rolland_freeform_2021}. Existing fabrication processes are used to produce freeform glass surfaces by sub-aperture material removal, and include computer-controlled polishing (CCP)  \cite{jones_optimization_1977}, magneto-rheological finishing (MRF) \cite{golini_magnetorheological_1999}, ion beam figuring\cite{weiser_ion_2009, schaefer_basics_2018}, reactive ion plasma figuring \cite{castelli_rapid_2012}. Freeform surfaces usually contain primarily low-spatial frequency aspheric departure, and the ideal fabrication process would accurately modify figure (low-spatial frequency height content) without introducing mid-spatial frequency (MSF) errors or roughness. 

Material removal processes generally do not achieve this ideal, and “mitigation or removal of MSFs is neither trivial nor universally demonstrated” \cite{rolland_freeform_2021}. As a simplification, the change in the optical surface from material removal processes is a convolution of the tool removal function and the dwell time \cite{jones_optimization_1977,wang_rifta_2020}. Small tools have a narrow tool removal function to enhance deterministic figuring performance but can introduce MSF error and require longer processing time. Large tools smooth out MSFs but usually degrade figure. Iteration between different processes is usually necessary to converge on a surface satisfying requirements over the all spatial frequencies \cite{suratwala_materials_2018}.

In contrast to material removal processes, stress figuring processes modify the spatially variant curvature of a mirror by applying deterministic stress to its substrate or back (non-optical) surface to impart a stress-induced bending moment. Modifying curvature, the second derivative of height, naturally suppresses MSF errors. In general, a surface has three curvature components, and to arbitrarily and accurately control the figure of a mirror over its full aperture requires controlling three plane stress components \cite{chalifoux_correcting_2018}. 
	
Researchers have long attempted to use a single stress component (equibiaxial stress, in which the stress state has two equal normal components) to correct figure errors, using thin films composed of metal \cite{yao_stress_2015}, dielectric\cite{yao_thermal_2019}, and piezoelectric \cite{patterson_ultralightweight_2013,deroo_deterministic_2018,kanno_development_2007,rodrigues_modular_2009} materials. For general figuring, however, equibiaxial stress leads to edge effects that limit either clear aperture or figuring accuracy \cite{vdovin_correction_2008}. 

These limitations are avoided, in theory, by controlling all three stress components. Stress generation processes capable of independently controlling three stress components include patterned thin films\cite{shen_stresses_1996,yao_stress_2021,zuo_experiments_2021}, ion implantation \cite{chalifoux_using_2017}, excimer laser writing\cite{beckmann_figure_2020}, patterned viscoelastic films\cite{beckmann_freeform_2022}, magnetostrictive films\cite{wang_deformation_2016}, and ultrafast laser writing \cite{chalifoux_figure_2021,graitzer_correcting_2011,seesselberg_optical_2019,bellouard_stress-state_2016}. So far, stress figuring has not been widely applied for creating freeform optics because no method has yet been shown to be accurate, stable, and scalable to high throughput or large optics. Iterative correction is critical for accuracy—to accommodate uncertainty in stress generation, substrate geometry and material properties—but only Yao, et al. \cite{yao_stress_2021} have so far demonstrated a multi-pass process. Yao, et al. also showed that all three stress components can be controlled if a single non-equibiaxial stress state is rotated to three orientations \cite{yao_stress_2021}, an insight that greatly simplifies using stress for mirror figuring by reducing the stress generation requirements to a single stress state. 
	
In this work, we present ultrafast laser stress figuring (ULSF, Section \ref{sec:ULSF}), a technique to accurately figure thin mirrors without creating MSF error. We describe systematic calibration and correction procedures (Section \ref{sec:MaterialsAndMethods}) that are applicable to a variety of other stress-generation methods. Finally, we demonstrate the accuracy, throughput, and stability of ULSF by figuring four exemplary thin mirrors (Section \ref{sec:ResultsAndDiscussion}). 

\section{Ultrafast laser stress figuring}
\label{sec:ULSF}
\subsection{ULSF process}
\label{ssec:ULSFProcess}
\begin{figure}[htbp]
\centering\includegraphics[width=13cm]{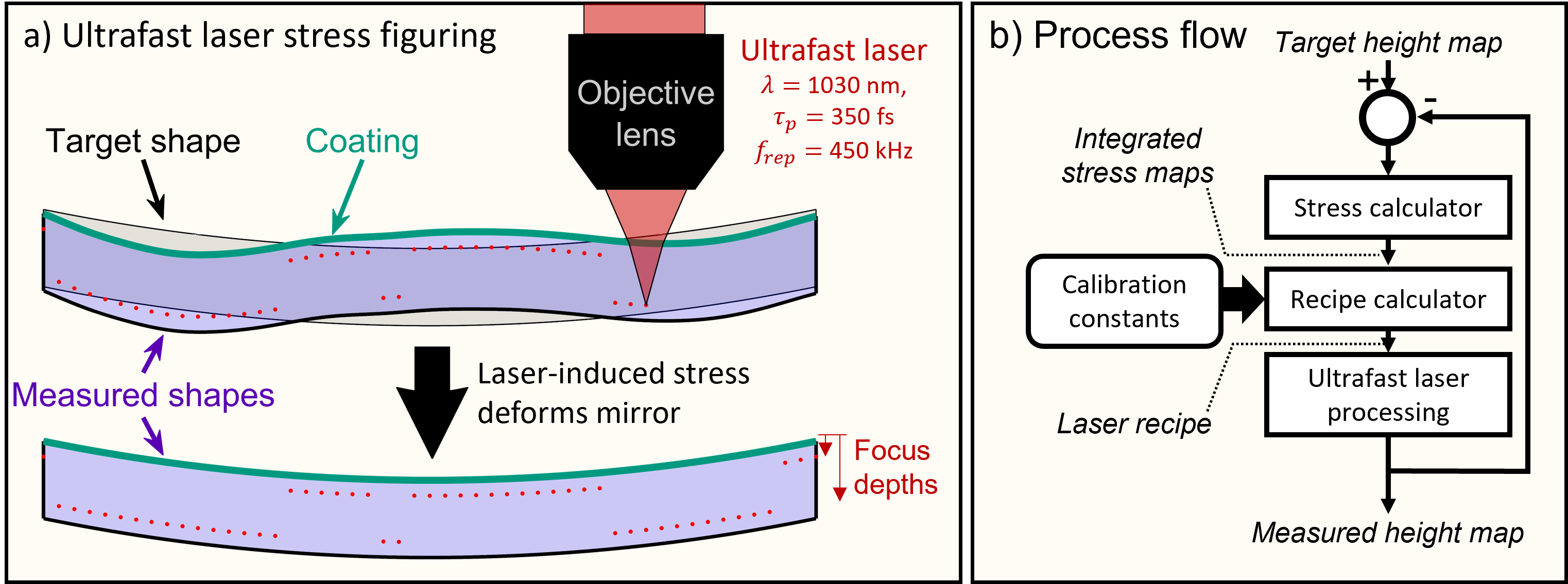}
\caption{a) Ultrafast laser stress figuring (ULSF) entails focusing an ultrafast laser at multiple depths in a mirror substrate, with coating on top (shown here) or bottom, depending on the coating. The stress induced by the laser spots bends the substrate to match the target shape. b) Process flow for applying ULSF to achieve a desired deformation of an optical surface. The stress calculator, recipe calculator, calibration procedure, and laser processing system are each described in the text.}
\label{fig:ULSFprocessFlow}
\end{figure}

Focusing ultrafast laser pulses, with sub-picosecond pulse duration, into fused silica generates the required non-equibiaxial stress state \cite{bellouard_stress-state_2016} for general figuring, and can be used iteratively\cite{zuo_ultrafast_2018}. In ULSF, we focus an ultrafast laser in a transparent substrate like fused silica, and we choose the 3-dimensional positions and laser parameters to generate a desired stress state at each point on the mirror (Fig. \ref{fig:ULSFprocessFlow}a). The stress bends the mirror from its original shape to a desired shape. ULSF can be applied with the mirror coating on top or bottom (both demonstrated in Section \ref{ssec:MirrorCorrection}). ULSF is non-contact and fluid-free, does not remove material or create particles, and does not require a special environment (e.g., vacuum). This work uses a fixed-position objective lens and translate the mirrors we correct, but mature laser processing technologies like scan mirrors, spatial-light modulators, or diffractive elements can dramatically increase throughput.

ULSF follows the process flow illustrated in Fig. \ref{fig:ULSFprocessFlow}b. We measure the initial mirror surface height map and subtract it from the target height map to obtain the desired deformation. In this paper the desired mirror prescription is flat, but in principle could be a freeform surface. We feed the desired deformation into a stress calculation routine to obtain maps of the three stress components. We then input the desired integrated stress maps, along with calibration constants, into recipe calculator software that allocates laser spots to achieve a desired laser spot density at each position and depth on the mirror. We implement this recipe on an ultrafast laser processing system, then measure the final mirror surface figure. We repeat this process until we satisfy a stopping condition: we meet the surface figure requirements, we reach the resolution limit of the correction process or metrology fixture, or we have insufficient space to write additional laser spots.

\subsection{Theoretical background}
\label{ssec:TheoreticalBackground}

When an ultrafast laser pulse is focused into a dielectric material, it is absorbed through nonlinear processes such as multiphoton absorption or avalanche ionization \cite{gattass_femtosecond_2008}. In fused silica, permanent strain is generated in the focal region, through structural change \cite{davis_writing_1996}, void formation \cite{glezer_ultrafast-laser_1997}, nanograting formation \cite{shimotsuma_self-organized_2003}, and thermal flow effects \cite{schaffer_bulk_2003}. The strained material, surrounded by a much larger substrate, develops stress that applies internal bending moments to the substrate. Ultrafast laser-induced stress has been measured in fused silica cantilevers, and was found to be tensile or compressive depending on laser pulse energy, repetition rate, polarization, or write speed \cite{bellouard_stress-state_2016,champion_direct_2012}.

A two-dimensional pattern of laser modifications written into a substrate will induce curvature akin to a stressed thin film. The integrated stress, or local stress integrated through the stressed region in the direction normal to the substrate surface, can be deduced from the curvature through Stoney’s equation\cite{stoney_tension_1909}. In general, there are three components of curvature ($\kappa_{xx},\kappa_{yy},\kappa_{xy}$), three components of integrated stress ($N_{xx},N_{yy},N_{xy}$), and the stressed film can be at any distance from the substrate midplane, $z_f$. The integrated stresses are related to the curvatures through a generalized Stoney’s equation adapted from Ref. \cite{suresh_thin_2009}:
\begin{align}
\label{eq:StressCurvature}
 \begin{bmatrix}
    N_{xx} \\
    N_{yy} \\
    N_{xy}
 \end{bmatrix}
 =\frac{E h_s^2}{6(1-\nu^2)} \frac{h_s/2}{z_f} 
 \begin{bmatrix}
    1 & \nu & 0\\
    \nu & 1 & 0\\
    0 & 0 & (1-\nu)
 \end{bmatrix}
 \begin{bmatrix}
    \kappa_{xx} \\
    \kappa_{yy} \\
    \kappa_{xy}
 \end{bmatrix}
\end{align}
where $E$ and $\nu$ are the elastic modulus and Poisson’s ratio of the substrate. In general, the curvatures and integrated stresses in Eq. \eqref{eq:StressCurvature} need not be uniform \cite{huang_anisotropic_2007}. The stressed film is usually located near the surface, so $z_f \approx \pm h_s/2$ .

The integrated stress required to deform a substrate with sinusoidal height amplitude $A$ depends on the spatial wavelength $\Lambda$. Consider a substrate with $\nu=0$ and a height error $w=A \sin{(2\pi/\Lambda)}$. From Eq. \eqref{eq:StressCurvature}, the required integrated stress is
\begin{align}
\label{eq:Nxx1D}
    N_{xx}=-\frac{E h_s^2}{6}\frac{h_s/2}{z_f} \left(\frac{2 \pi}{\Lambda}\right)^2 A \sin{\left(\frac{2\pi x}{\Lambda}\right)}
\end{align}
and in this specific case the other two stress components are zero. Since the stress applied by any stress-generation method has limited magnitude, stress-based figuring is best suited to low-spatial frequency errors, thin substrates, or small displacements. This also suggests that mid-spatial frequency errors are difficult to unintentionally introduce by errors in $N_{xx}$.

Stress figuring, when used to introduce low-spatial frequency deformations, has low sensitivity to lateral or depth position errors of the integrated stress. The relative error in the displacement amplitude $\delta A / A$ depends on the lateral position error $\delta x$ and the depth error $\delta z_f$. A small lateral position error produces an integrated stress error $\delta N_{xx}=(\partial N_{xx} / \partial x)\delta x$. As detailed in Supplement 1, the relative RMS amplitude error (RMS values denoted by $\langle\bullet\rangle$) that results from each position error is
\begin{align}
\label{eq: sensitivities}
    \langle{\frac{\delta A}{A}}\rangle_{\delta zf=0} = \langle{\frac{2\pi \delta x}{\Lambda}}\rangle, &&
    \langle{\frac{\delta A}{A}}\rangle_{\delta x=0} = \langle{\frac{\delta z_f}{z_f}}\rangle.
\end{align}

For a substrate with $\delta x=15\ \mu m,\Lambda=10\ mm,\delta z_f=2.5\ \mu m,z_f=0.5\ mm$, the relative amplitude error is <2\%. These tolerances are readily achievable with commercial positioning stages. Mirror substrate thickness variation will lead to depth error, but several microns is within commercial tolerances. The effects of these errors are mitigated by iterative figure correction.

The ideal stress-generation method would apply large-magnitude stress with independent and accurate control of the three components. Yao et al.\cite{yao_stress_2021} showed that independent control of the three stress components can be accomplished by rotating one base stress state to three orientations, as long as the base stress state is non-equibiaxial (i.e., $N_{xx}\neq N_{yy}$). We have a base stress state oriented at $\phi_0=0^\circ, N_{xx}^0,N_{yy}^0,N_{xy}^0$, which we rotate to orientations $\phi=0^\circ,120^\circ,240^\circ$ and multiply by constants $a_1,a_2,a_3$. We represent the integrated stress state as $\vec{N}=\mathbf{T} \vec{a}$ (derived in Supplement 1),
\begin{align}
    \begin{bmatrix}
        N_{xx}\\N_{yy}\\N_{xy}
    \end{bmatrix}=
     \begin{bmatrix}
       N_{xx}^0 && \frac{N_{xx}^0+3N_{yy}^0}{4} + \frac{\sqrt{3} N_{xy}^0}{2}&& \frac{N_{xx}^0+3N_{yy}^0}{4} - \frac{\sqrt{3} N_{xy}^0}{2}\\
       N_{yy}^0 && \frac{3 N_{xx}^0+N_{yy}^0}{4} - \frac{\sqrt{3} N_{xy}^0}{2}&& \frac{3 N_{xx}^0+N_{yy}^0}{4} + \frac{\sqrt{3} N_{xy}^0}{2}\\
       N_{xy}^0 && -\frac{\sqrt{3}}{2}(N_{xx}^0-N_{yy}^0) + \frac{N_{xy}^0}{4}&& \frac{\sqrt{3}}{2}(N_{xx}^0-N_{yy}^0) + \frac{N_{xy}^0}{4}
     \end{bmatrix}
     \begin{bmatrix}
        a_0\\a_1\\a_2
     \end{bmatrix}.
\end{align}
Achieving a given stress state $\vec{N}$ requires magnitudes $\vec{a}=\mathbf{T}^{-1} \vec{N}$. The components of $\vec{a}$ may be positive or negative, and $\det{\mathbf{T}}\neq 0$ if and only if $N_{xx}^0\neq\pm N_{yy}^0$. Therefore, achieving any stress state by rotating a base stress state to three orientations is only possible if $N_{xx}^0\neq\pm N_{yy}^0$ and if we can control the components of $\vec{a}$ to be positive or negative.

For ULSF, we fix the laser pulse energy and repetition rate, and vary the areal density of laser spots, $D$, and the distance from the midplane, $z_f$. Varying $D$ alone varies the magnitude of $\vec{a}$ components without changing their sign, and changing the sign of $z_f$ (i.e., writing above and below the midplane) changes the sign of the $\vec{a}$ components. Following \cite{bellouard_stress-state_2016}, in this work we achieved the required non-equibiaxial base stress state by orienting the laser electric field polarization to be normal to the writing direction.  We wrote spots above and below the midplane, at 3 orientations. We perform a calibration (Section \ref{ssec:CalibrationConstants}) so we do not require exactly the same stress state in all three orientations.

\section{Materials and methods}
\label{sec:MaterialsAndMethods}
\subsection{Laser processing equipment}
\label{ssec:LaserProcessingEquipment}
\begin{figure}[htbp]
\centering\includegraphics[width=7cm]{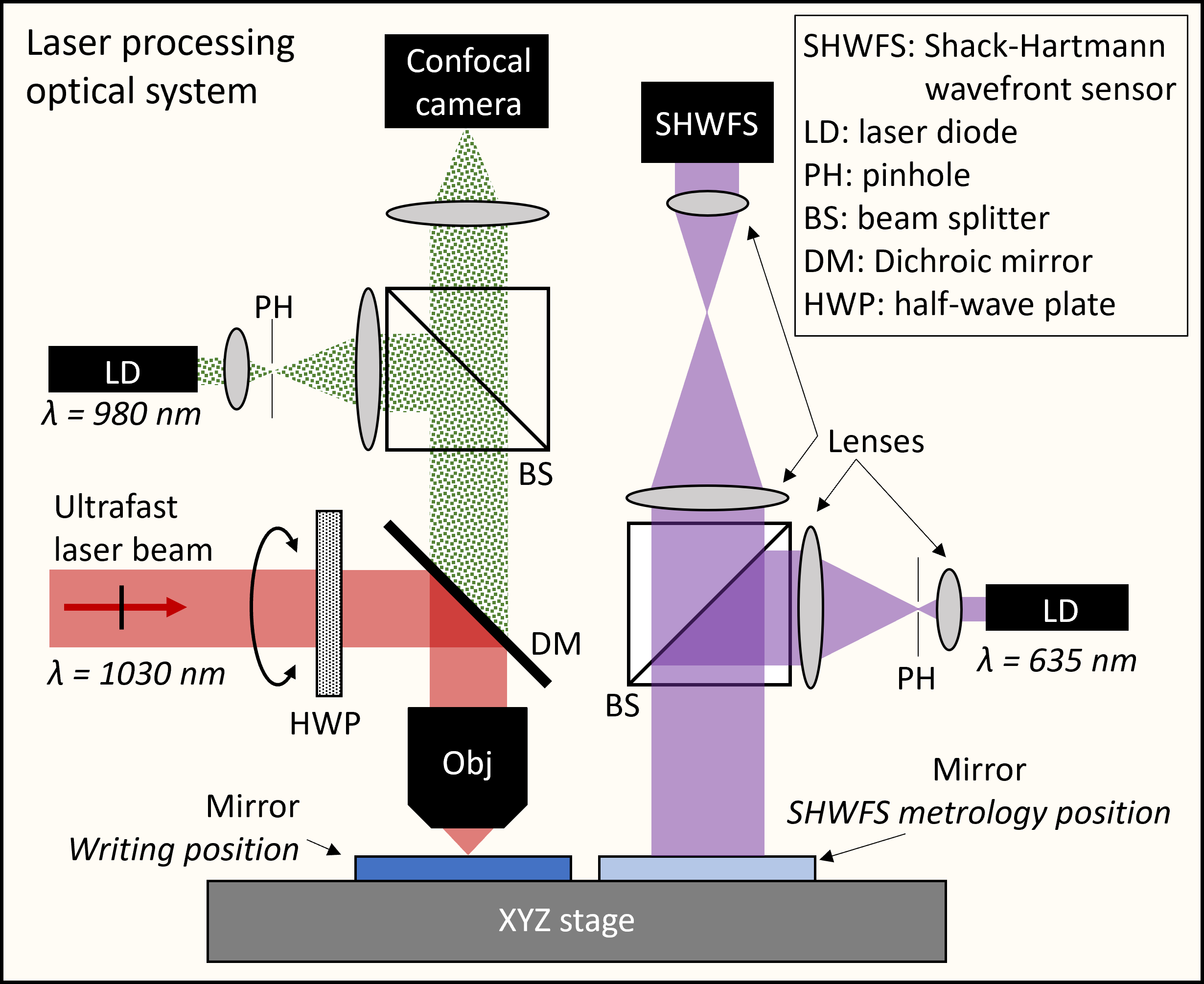}
\caption{Laser processing optical system. The ultrafast laser beam, with its electric field initially vertical, passes through a rotatable halfwave plate (HWP), reflects off a dichroic mirror (DM), then focuses through an objective lens (Obj, 0.4 NA) into the Mirror. A laser diode (LD) with $\lambda$=980 nm illuminates a pinhole (PH) that is imaged onto a CMOS camera when a surface of the Mirror is positioned at the objective focal plane. The Mirror is moved to the Shack-Hartmann wavefront sensor (SHWFS) position for in-situ surface measurement. The SHWFS system incorporates an afocal telescope to magnify the Mirror surface slopes on the SHWFS camera.}
\label{fig:LaserSystemOptics}
\end{figure}

We used an industrial ultrafast laser (Trumpf TruMicro 2030) with $\lambda$ = 1030 nm, pulse duration 350 fs, repetition rate 450 kHz, pulse energy $E_p$=650 nJ. We rotated its polarization using a half-waveplate on a motorized rotation stage, expanded its beam to 7 mm diameter, and focused it through an objective lens with 0.4 NA (ThorLabs LMH-20X-1064). The mirrors were mounted on a 3-axis stage system (Aerotech PlanarDL-200XY and AVSI100-25) and translated at a constant speed of 45 mm/second. The stage has position-sensitive electrical signaling to trigger single laser pulses with sub-$\mu$m resolution. The system is illustrated in Fig. \ref{fig:LaserSystemOptics}.

\subsection{Mirrors}
\label{ssec:Mirrors}
We applied ULSF to fused silica mirrors with various geometry and coatings (Table \ref{tab:MirrorParameters}). We performed calibration on a 25.4 mm-diameter mirror (CAL, Section \ref{ssec:CalibrationConstants}) with a 10 mm-long edge flat, and we corrected four 100 mm-diameter mirrors (M1-M4, Section \ref{ssec:MirrorCorrection}) that had a 32.5 mm flat. M1 and M2 were double-side polished to 20-10 scratch-dig surface quality, and all others were polished to 60-40 scratch-dig. The mirrors were coated on one side with an aluminum or dielectric coating. Aluminum (100 nm) with a SiO$_2$ overcoat (200 nm) was deposited via electron-beam evaporation with the substrate at room temperature. The dielectric coating was a 3-bilayer Bragg coating (70.6 nm TiO2, 112 nm SiO2) tuned to 85\% reflectivity at 633 nm and 6\% reflectivity at 1030 nm, and was deposited via ion-assisted thermal evaporation with the mirror substrates heated to 100 $^\circ$C. The clear aperture of each mirror was limited by the coating coverage and edge roll-off of the mirror surfaces.

\begin{table}
\centering
    \begin{threeparttable}
        \caption{Mirror parameters}
        \label{tab:MirrorParameters}
            \begin{tabular}{m{3cm}| m{0.8cm} m{0.8cm} m{0.8cm} m{0.8cm} m{0.8cm}}
                Mirror&	CAL&	M1&	M2&	M3&	M4\\
                \hline
                Thickness [mm]&	1&	1&	1&	1&	2\\
                Diameter [mm]&	25.4&	100&	100&	100&	100\\
                Clear aperture [mm]&	22.9&	97&	97&	95&	95\\
                Material$^a$&	C&	H&	H&	H&	C\\
                Coating$^a$&	D&	 D&	D&	A&	A\\
                Laser side& Top&	Top&	Top& Bot.&	Bot.\\
                \hline
            \end{tabular}
        \begin{tablenotes}
            \small
            \item $^a$ C – Corning 7980; H – Heraeus Suprasil 313; D – dielectric, 6-layer Bragg coating; A – aluminum/SiO$_2$ coating
        \end{tablenotes}
    \end{threeparttable}
\end{table}

\subsection{Surface metrology and stress analysis}
\label{ssec:SurfaceMetrology}
We measured mirror surface height off-machine using a Fizeau interferometer ({\"A}pre Instruments SR100$\mid$HR), or on-machine using a Shack-Hartmann wavefront sensor (SHWFS, Fig. \ref{fig:LaserSystemOptics}). For interferometer measurements, we measured (with 10-measurement averages) the surface of each mirror before and after writing a laser spot recipe into it. We fit up to 6$^{th}$ order normalized Zernike polynomials ($Z_n^m$ with order $n\leq 6$ and frequency $|m| \leq$6) to the measured mirror surfaces. We removed tilt ($Z_1^{\pm 1}$) and piston ($Z_0^0$) from all surface maps. Our mirror metrology mount holds the mirrors vertically, located by three cylinders below and on the side, and three point-contacts behind the mirror. When measuring mirrors with 100 mm diameter and 1 mm thickness, we additionally averaged five 10-average measurements, re-mounting the mirror each time, to reduce uncertainty from mounting-induced deformation. The metrology noise spectrum for 1 mm-thick mirrors is included as a dashed line in Fig. \ref{fig:ZernSpectra}.

We used on-machine metrology for scans of depth and line-spacing (Section \ref{ssec:LinearityTests}). The SHWFS system (Fig. \ref{fig:LaserSystemOptics}) comprises a Shack-Hartmann wavefront sensor (ThorLabs WFS40-14AR), laser diode illumination ($\lambda$=635 nm), and an afocal telescope with 22 mm aperture. We fit power ($Z_2^0$) and astigmatism ($Z_2^{\pm 2}$) terms to the measured surface slopes, and calibrated this system against the Fizeau interferometer.

We calculate the integrated stress analytically, from the measured surface maps and substrate geometry, using Stress Field II from Chalifoux et al.\cite{chalifoux_correcting_2018}. For convenience, in this work we always calculate the integrated stress with $z_f=h_s /2$, then when we allocate laser spots (Section \ref{ssec:SpotAllocation}) we apply a scaling factor $\eta=t_f/(h_s /2)$ to our calibration constants to account for the depth of each laser spot.
We measured height errors with spatial wavelengths from about 0.05 mm to 1.3 mm using a white light interferometer (WLI, Zygo NewView 8800) with a 10$\times$ Mirau objective and 0.5$\times$ zoom (1.3 mm field of view). We measured the coated surface of each mirror at 25 locations (on a 5$\times$5 square grid with 15 mm spacing) before and after correction. We chose a wide field of view to detect potential print-through of the 0.5-mm triangular pattern (Fig. \ref{fig:DICimage}a) into the coated surface, which might occur from damage to the coating or bulging of the fused silica.

\subsection{Spot allocation}
\label{ssec:SpotAllocation}
When correcting mirror figure, we first calculate the integrated stress maps at a grid of points on the mirror. We then allocate spots (precisely defined in Section \ref{ssec:CalibrationConstants}), each of which generates some amount of stress, to achieve this stress state. We use our measured calibration matrix $\mathbf{C}$ (a 3$\times$6 matrix, Section \ref{ssec:CalibrationConstants}), containing constants $C_{sp}$, to allocate laser spots into each of 6 paths (index $p$=P1-P6) and achieve the desired stress state ($s=N_{xx},N_{yy},N_{xy}$) at each grid point. Three paths lie below the substrate midplane ($z_f<0$ for $p$=P1-P3), and three paths lie above it ($z_f>0$ for $p$=P4-P6). The calibration constants for the deep paths approximately mirror those of the shallow paths but with opposite signs. 

We use a simple iterative algorithm to determine the spot density, $\vec{D}$ (a 6$\times$1 vector), in each path at each grid point that achieves the desired stress, $\vec{N}$ (a 3$\times$1 vector), at that point. To start the spot allocation process, we arbitrarily extract the first three columns of the calibration matrix, invert it and multiply by the desired stress state to determine the spot density in P1-P3, $\vec{D}_{123}(x,y)=\mathbf{C}_{123}^{-1} \vec{N}(x,y)$. Typically, this will result in some negative values of $\vec{D}_{123}$, which is a non-physical result. For those negative elements, we switch from the deep to the shallow path and re-calculate $\vec{D}$. We repeat this until all values of $\vec{D}$ are positive or zero at all grid points (typically 1-3 iterations). Once the spot density $\vec{D}$ is calculated, we proceed to allocate spots to specific locations. The spots are arranged into lines at three orientations, forming triangular cells surrounding a grid point (Fig. \ref{fig:DICimage}a). We add spots to each cell until it has no additional space, and we then begin adding spots to a shallower layer for that cell. We scale the integrated stress contributed by spots at different depths.

Iterative correction is usually needed to achieve a desired figuring accuracy. For subsequent corrections, we load a previous spot allocation and continue allocating spots to avoid writing in locations where we already wrote. For very small deformations (< 100 nm RMS), correction accuracy is limited by only being able to write an integer number of spots. Power ($Z_2^0$) and astigmatism ($Z_2^{\pm 2}$) terms are particularly sensitive to the mean error in each integrated stress component across the mirror. To compensate for this error for those terms, for each grid point we calculate the error between the desired integrated stress $\vec{N}$ and the expected integrated stress based on the allocated spot density $\vec{D}_{allocated}$ resulting from the integer number of spots and their depth, $\vec{N}_{error}=\mathbf{C} \vec{D}_{allocated} \vec{N}$. We average $\vec{N}_{error}$ over all grid points, and for corrections after the first pass, typically resulting in around $\pm$1 N/m error in each stress component. For a 1 mm-thick 100 mm-diameter fused silica mirror, this would result in an unacceptable RMS error > 20 nm RMS for power and astigmatism. We reduce this error to < 0.05 N/m by increasing the grid point spacing to expand the area for each grid point, or by iteratively adding a small uniform integrated stress (which may be different for each of the three components) to the desired stress map.

\section{Experimental results and discussion}
\label{sec:ResultsAndDiscussion}
\subsection{Calibration constants}
\label{ssec:CalibrationConstants}
Our correction strategy relies on writing a laser spot pattern at three orientations, both above and below the substrate midplane, thus requiring six calibration paths. Within each path, we write a fixed number of spots surrounding each grid point. We divide the integrated stress generated by each calibration path (see Dataset 1\cite{chalifoux_arnold_laverty_2022} for surface measurements and spot allocations) by the areal density of the pattern ($D$=185 mm$^{-2}$) to obtain our calibration matrix (Table \ref{tab:CalibrationConstants}).

\begin{table}
\centering
    \begin{threeparttable}
        \caption{Calibration constants, $C_{sp}$ [N/m/mm$^{-2}$]}
        \label{tab:CalibrationConstants}
            \begin{tabular}{m{1.2cm}| m{1cm} m{1cm} m{1cm} m{1cm} m{1cm} m{1cm}}
                Stress  & \multicolumn{6}{c}{Path$^a$}\\
                \cline{2-7}
                comp. $s$ & P1&	P2&	P3&	P4&	P5& P6\\
                \hline
                $N_{xx}$&	0.622&	1.876&	2.101&	-0.430&	-1.289&	-1.297\\
                $N_{yy}$&	2.249&	1.045&	0.916&	-1.445&	-0.711&	-0.520\\
                $N_{xy}$&	0.192&	-0.712&	0.644&	-0.101&	0.464&	-0.393\\
                \hline
            \end{tabular}
        \begin{tablenotes}
            \small
            \item $^a$ $z_f$=-394 $\mu$m (P1-P3) and +206 $\mu$m (P4-P6)
        \end{tablenotes}
    \end{threeparttable}
\end{table}

\begin{figure}[htbp]
\centering\includegraphics[width=13cm]{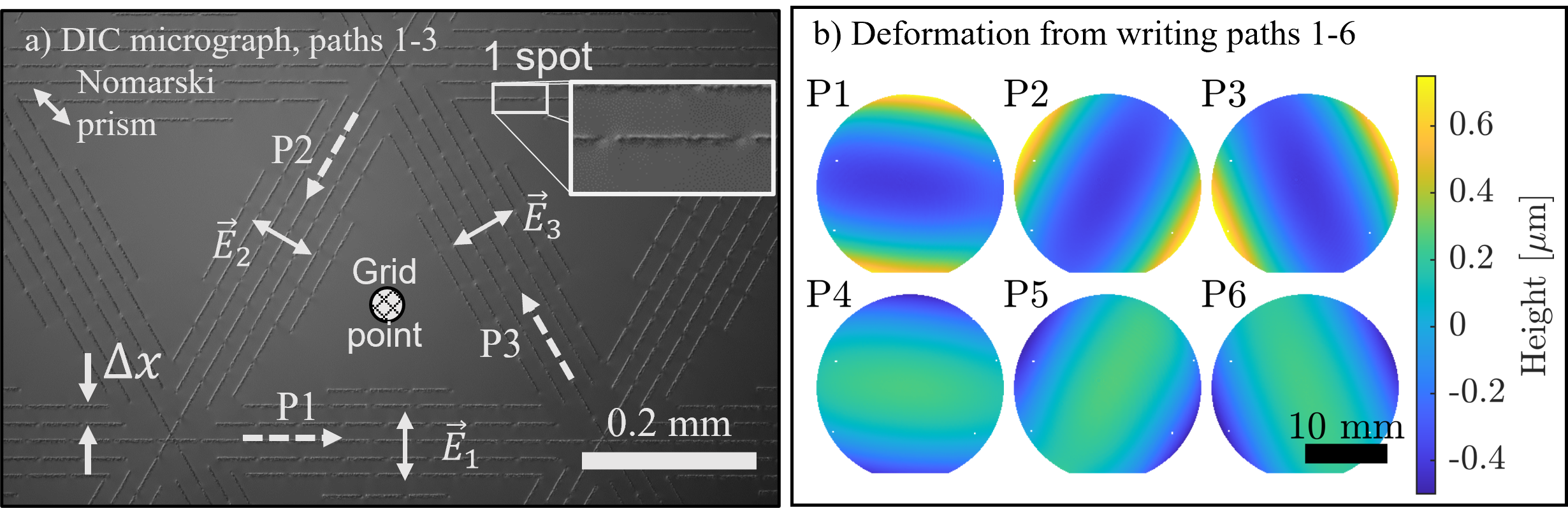}
\caption{Laser-written modifications at 3 orientations for CAL. a) a differential-interference contrast (DIC) micrograph, taken using a 550 nm interference filter at 20$\times$ magnification (0.4 NA) of laser-written modifications surrounding a grid point. Each path (P1-P3 are shown here, P4-P6 are written at a different depth) is written along a different direction, with the electric field $\vec{E}$ perpendicular to the writing direction. A laser spot is a segment 45 $\mu$m long made by firing 450 laser pulses while traveling , and the lines are spaced by a distance $\Delta x$. b) Writing each path P1-P6 causes deformation, primarily power ($Z_2^0$) and astigmatism ($Z_2^{\pm2}$). P1-P3 are deep in the substrate and exhibit larger deformation than P4-P6 proportional to their distance from the midplane $z_f$.}
\label{fig:DICimage}
\end{figure}

Here we define a spot as a line segment 45 $\mu$m long (Fig. \ref{fig:DICimage}a) that comprises 450 laser pulses fired while the substrate is translated at 45 mm/second along one of 3 directions, 0$^\circ$, 120$^\circ$, or 240$^\circ$.  The distance between spots is 50 $\mu$m and the laser polarization (E-field) is oriented perpendicular to the line. Lines are spaced by $\Delta x=20\ \mu m$. Before and after writing each path, we measure the surface using the Fizeau interferometer. The deformations induced by each path (Fig. \ref{fig:DICimage}b) are predominantly power ($Z_2^0$) and astigmatism ($Z_2^{\pm2}$), from which we calculate the unform stress components created by each path, using Eq. \eqref{eq:StressCurvature}. We used one calibration mirror (CAL) with coating on top for M1 and M2, and another with coating on bottom for M3 and M4 (with similar constants). The distance from the midplane $z_f$ is calculated using the surface positions measured with the confocal imaging system (Fig. \ref{fig:LaserSystemOptics}). The distance $z_f$ is not used when calculating the calibration constants, but it is used when scaling to different layer depths during recipe calculation.

\subsection{Linearity tests}
\label{ssec:LinearityTests}
\begin{figure}[htbp]
\centering\includegraphics[width=13cm]{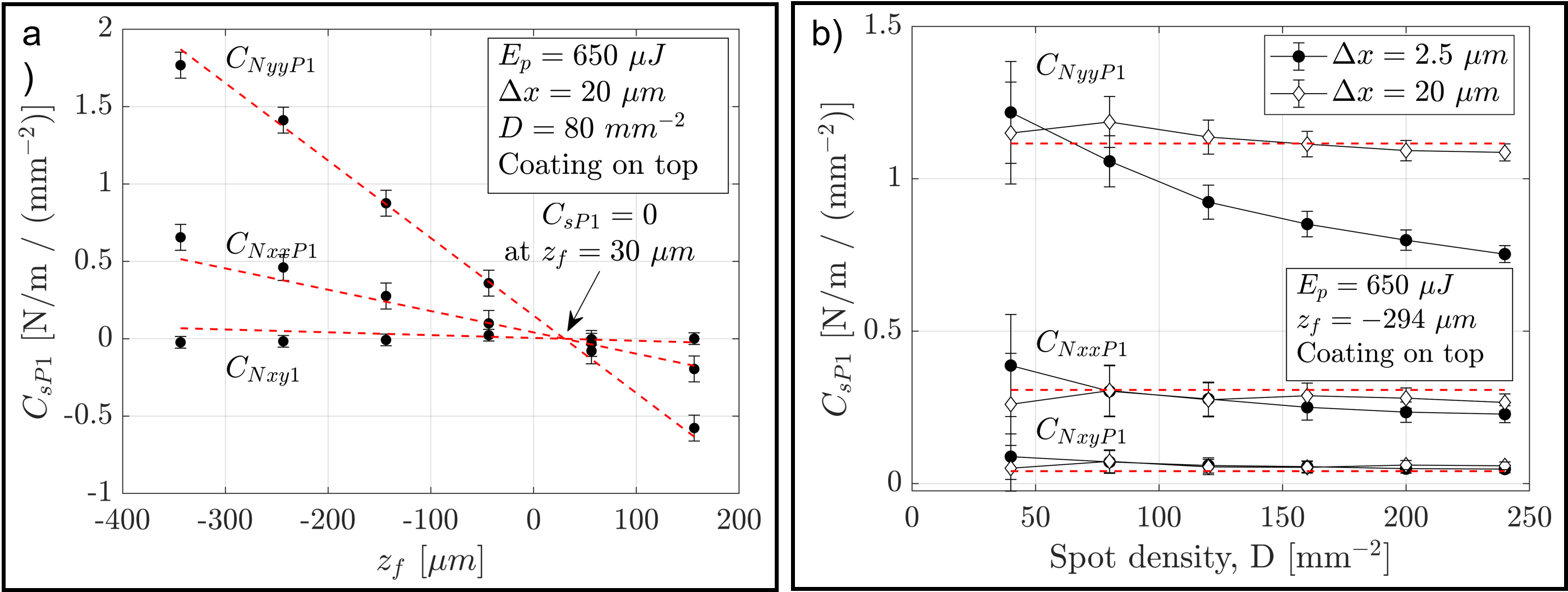}
\caption{Calibration constants measured under different conditions. The error bars ($\pm 1 \sigma$) are based on the SHWFS measurement repeatability. The dashed lines are the expected values based on Eq. \eqref{eq:StressCurvature} and the calibration constants, with a 30 $\mu$m depth offset applied. a) Varying distance from the midplane $z_f$ results in a linear variation of $C_{sp}$, consistent with Eq. \eqref{eq:StressCurvature}. b) Spacing lines with $\Delta x=20\ \mu m$ results in more consistent calibration constants than $\Delta x=2.5\ \mu m$.}
\label{fig:CalLinearity}
\end{figure}

We used the SHWFS (Fig. \ref{fig:LaserSystemOptics}) to measure the variation in calibration constants $C_{sp}$ for 3 samples under varying conditions: distance to the midplane $z_f$ (Fig. \ref{fig:CalLinearity}a), spot density $D$, and line spacing $\Delta x$ (Fig. \ref{fig:CalLinearity}b). Since we always calculate integrated stress assuming $z_f=h_s /2$, the calibration constants ideally vary linearly with $z_f$ and do not vary with $D$. We compared our measurements with the expected calibration constants based on Eq.\eqref{eq:StressCurvature} using the values in Table \ref{tab:CalibrationConstants} (dashed lines in Fig. \ref{fig:CalLinearity}). We found good agreement when the effective depth of the stressed layer is 30 $\mu$m deeper than our estimated depth (red dashed line, Fig. \ref{fig:CalLinearity}a). This offset might be due to systematic error in our depth estimate (based on stage motion and our confocal microscope, see Fig. \ref{fig:LaserSystemOptics}), or it may occur if the tip of the laser-induced modifications does not exactly coincide with the center of stress. Regardless of the cause, we apply this calibrated offset when allocating spots for correcting mirrors (Section \ref{ssec:MirrorCorrection}). Our spot allocation procedure assumes the calibration constants are independent of spot density D. We found that spacing lines farther apart reduces the dependence of the calibration constants on spot density (Fig. \ref{fig:CalLinearity}b), so we used $\Delta x=20 \mu m$ for  mirror correction (Section \ref{ssec:MirrorCorrection}). For $\Delta x=2.5 \mu m$, the interacting stress fields from adjacent lines may explain the diminished stress generated per spot. 

\subsection{Mirror correction}
\label{ssec:MirrorCorrection}
We corrected four fused silica mirrors, with different combinations of coating and thickness (Table \ref{tab:MirrorParameters}), to demonstrate the versatility of ULSF. We aimed to correct all terms up to 6$^{th}$ order Zernike polynomials for each mirror by writing spots following our spot allocation procedure (see Fig. S1 for integrated stress fields and corresponding spot densities), and applying 2-3 correction passes for each mirror. To detect any relaxation of the imparted stress, we measured each mirror for 3-5 weeks (we refer to this as 1 month for brevity) at approximately 1-week intervals after correction. Here we primarily report on M1, but the results for M2-M4 are similar except where noted (see Supplement 1).

\begin{figure}[htbp]
\centering\includegraphics[width=13cm]{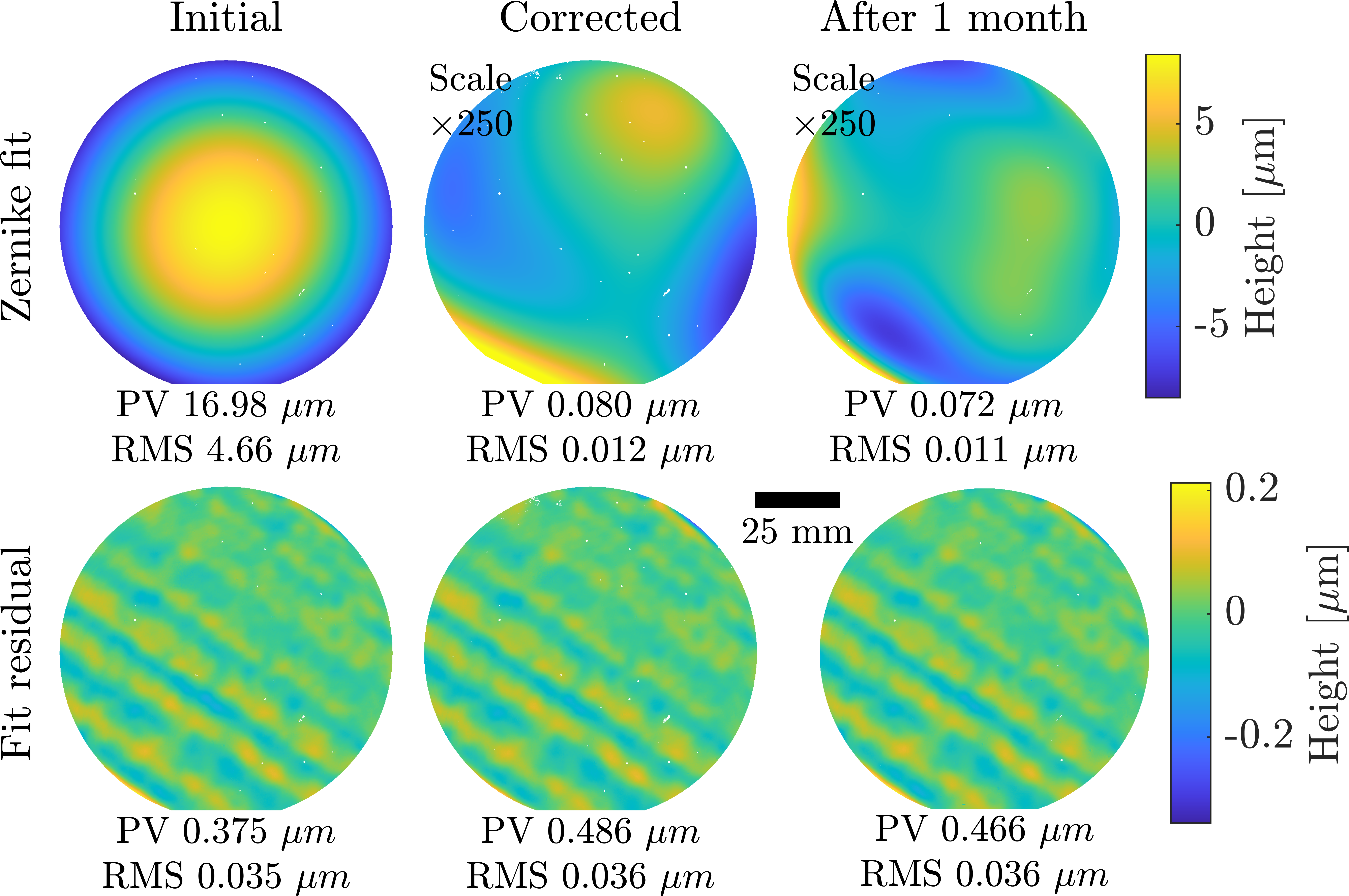}
\caption{Surface height maps of M1 before and after correction, and 1-month post-correction. We aimed to reduce the 6$^{th}$ order Zernike fit (top) to zero while leaving the fit residual (bottom) unaffected. The fit residual represents pre-existing MSF errors, which ULSF does not affect. All plots only include data from the 97 mm clear aperture of M1.}
\label{fig:Mirror1HeightMaps}
\end{figure}

We separate the measured surface height map into two components: the Zernike fit up to 6$^{th}$ order, and the residual of the surface map after subtracting the 6$^{th}$ order Zernike fit. The Zernike fit ideally reduces to zero and represents the accuracy of correction, whereas the residual should remain unchanged. Table \ref{tab:FizeauResults} lists the RMS surface height of each mirror after each correction pass and after \textasciitilde 1 month post-correction. M1 and M2 are stable but M3 and M4 change after correction, which we discuss shortly. The height maps for M1 (Fig. \ref{fig:Mirror1HeightMaps}) show that the Zernike fit is driven to 12 nm RMS over a 97 mm clear aperture, while the fit residual is not significantly affected. Similar results were achieved for M2-M4 (Fig. S2).

\begin{table}
\centering
    \caption{Surface height measurements [$\mu$m RMS] over clear aperture}
    \label{tab:FizeauResults}
        \begin{tabular}{m{2cm}| m{1.5cm}| m{1cm} m{1cm} m{1cm} m{1cm}}
            \multicolumn{2}{c}{} & M1&	M2&	M3&	M4\\
            \hline
            \multirow{5}{2cm}{Zernike fit}& Init.&    4.656&	4.616&  0.746&  1.382\\
            &                             Pass 1&   0.0614&	0.306&	0.193&	0.176\\
            &                             Pass 2&   0.0123&	0.0703&	0.0300&	0.0224\\
            &                             Pass 3&   N/A&	0.0193&	0.0135&	0.0116\\
            &                             1 month&    0.0105&	0.0212&	0.0960&	0.0723\\
            \hline
            \multirow{3}{2cm}{Fit residual} & Init.&  0.0351& 0.0902& 0.1561& 0.0312\\
            &                               Pass 3& 0.0355 & 0.0845& 0.1549& 0.0305\\
            &                               1 month&  0.0359& 0.0843& 0.1560& 0.0331\\
            \hline
        \end{tabular}
\end{table}

\begin{figure}[htbp]
\begin{subfigure}{0.495\textwidth}
\centering\includegraphics[width=1\linewidth]{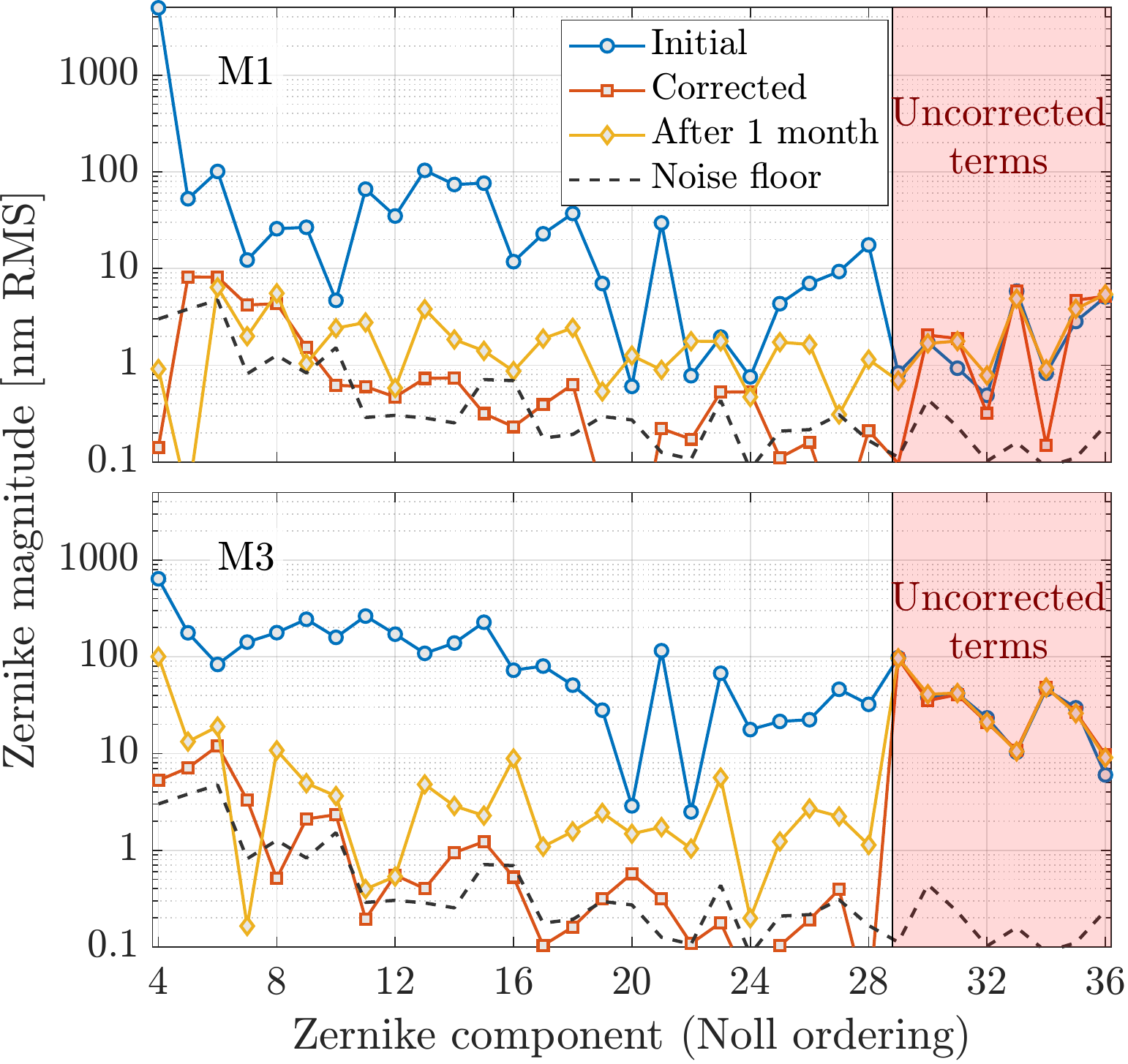}
\end{subfigure}
\begin{subfigure}{0.495\textwidth}
\centering\includegraphics[width=1\linewidth]{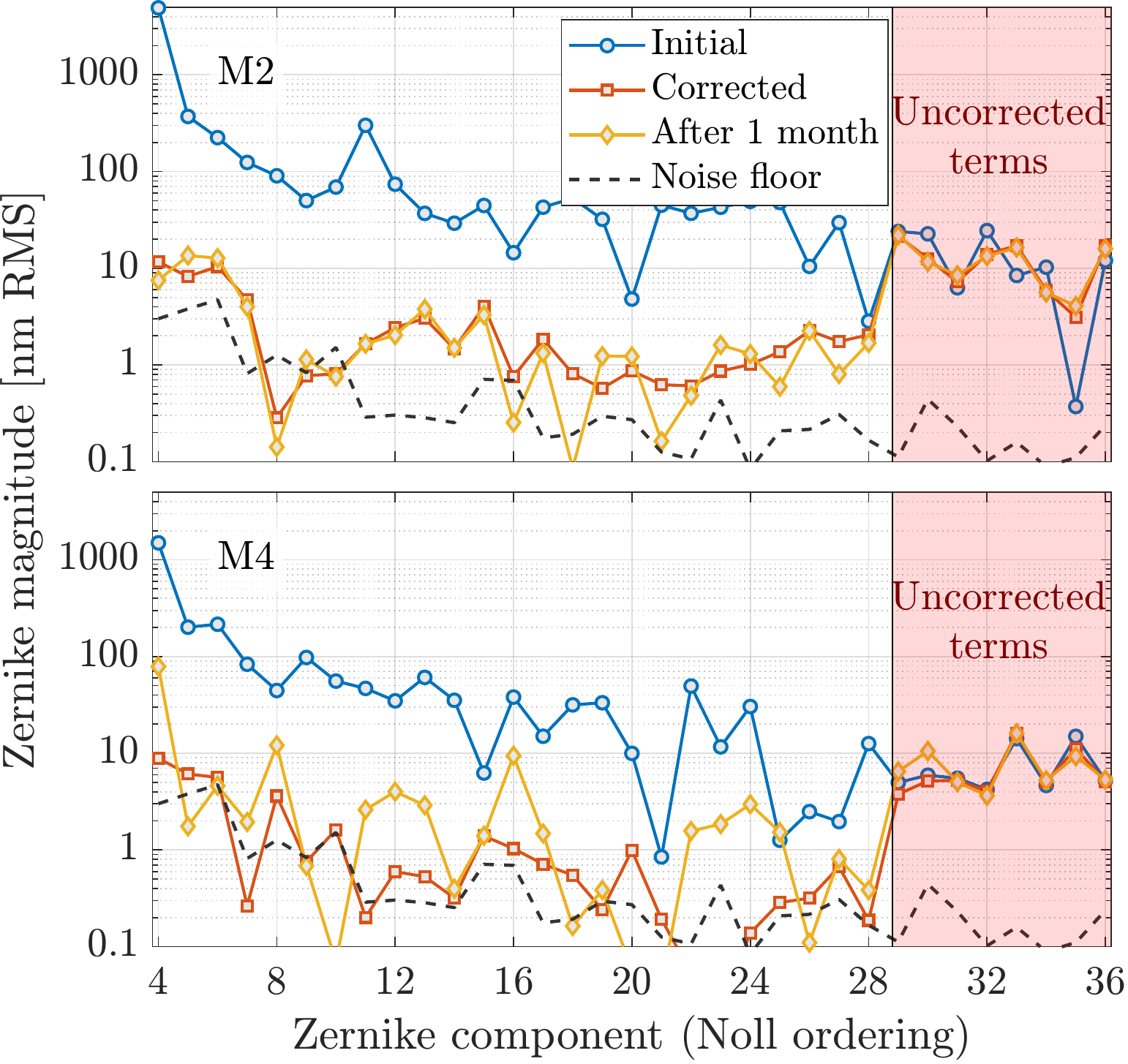}
\end{subfigure}
\caption{Zernike spectra of M1-M4 before and after correction, and after 1 month. We aimed to correct up to term 28 and leave higher-order terms unaffected (highlighted in red). The Zernike coefficients are ordered according to Noll \cite{noll_zernike_1976}. See text for measurement noise calculation.}
\label{fig:ZernSpectra}
\end{figure}

The Zernike spectra of M1-M4 are plotted on a log scale in Fig. \ref{fig:ZernSpectra}. The noise floor (Fig. \ref{fig:ZernSpectra}, black dashed line) was calculated for the 1 mm-thick mirrors from the standard deviation of each Zernike component from each of the 5-measurement sets for those mirrors. The noise spectrum is the mean of those standard deviations divided by $\sqrt{5}$ to reflect that our reported spectra are the mean of 5-measurement sets. The noise spectrum does not account for day-to-day variations. The Zernike spectra reveal that we successfully corrected the first 6 orders (terms 4-28) nearly to the noise floor for all 4 mirrors, whereas the 7$^{th}$ order polynomials (terms >28) remain nearly the same.

\begin{figure}[htbp]
\centering\includegraphics[width=7cm]{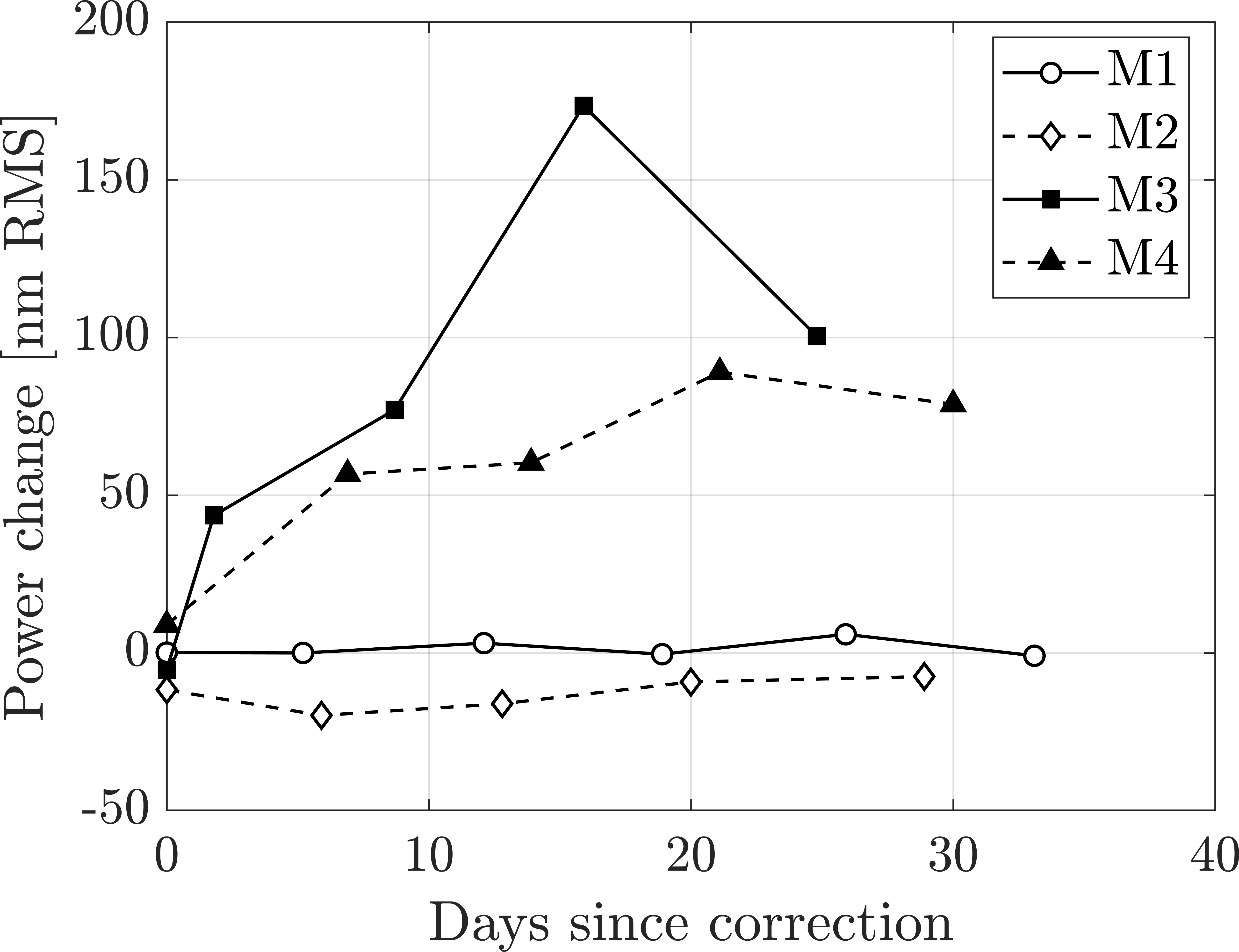}
\caption{The power term of each mirror M1-M4 for about a month following correction. M1 and M2, which are coated with multilayer dielectric films, appear stable. M3 and M4, which are coated with aluminum films, change. M3 changed at a similar rate over the 6 months prior to correction.}
\label{fig:PowerChangeOverTime}
\end{figure}

After 1 month, the Zernike spectra of M1 and M2 remain near the noise floor. These mirrors are stable, unlike M3 and M4. The power terms ($Z_2^0$, term 4 in Fig. \ref{fig:ZernSpectra}) of M3 and M4 change over time (Fig. \ref{fig:PowerChangeOverTime}). The distinguishing characteristic between M1/M2 and M3/M4 is that M3 and M4 were coated with 300 nm Al/SiO$_2$. M3 was coated 6 months prior to figure correction, and comparing post-coating to pre-correction measurements reveals a change in power of approximately 1300 nm RMS, or 6.3 nm/day. The post-correction change of M1 is around 4-6 nm/day. Post-coating measurements of M1, M2, and M4 are unavailable. The change we observe in M3 could be caused by stress relaxation in the films\cite{scherer_optical_1996} or humidity, but we cannot determine from this data whether the laser processing affects these films. Evaluating stability of these films after laser processing is under investigation, and will require producing initially-stable metal films.

\begin{figure}[htbp]
\centering\includegraphics[width=7cm]{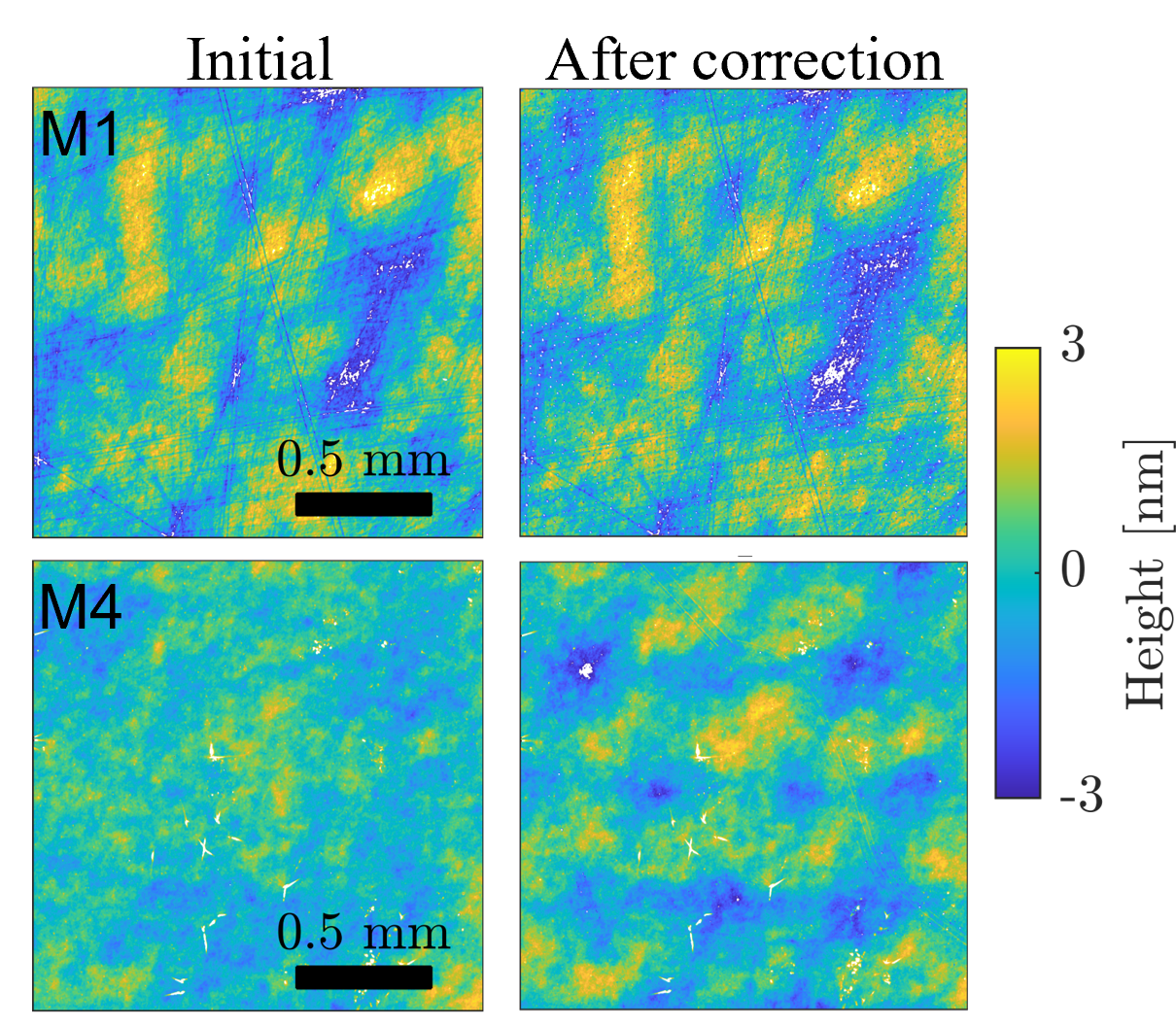}
\caption{WLI measurements of the center of M1 and M4. M1 (like M2 and M3) does not show any pattern print-through after coating. Some locations on M4 have a visible hexagonal grid with 0.5 mm spacing, as shown here. }
\label{fig:WLImaps}
\end{figure}

We took WLI measurements before and after correction at 25 locations on each mirror to detect print-through of our 0.5 mm triangular pattern (available in Dataset 1\cite{chalifoux_arnold_laverty_2022}). Table \ref{tab:WLIResults} shows the RMS height before and after correction at 5 locations (center and each corner of the pattern, respectively) on each mirror, and the mean change in RMS height of all 25 locations. Figure \ref{fig:WLImaps} shows the WLI measurements at the center of two mirrors: M1 and M4. M1-M3 show only angstrom-level RMS height changes, with no obvious print-through. However, M4 exhibits print-through with amplitude up to 3 nm at 9 of the 25 locations we measured. Figure \ref{fig:WLImaps} displays one of the worst locations on M4. For all mirrors, we kept the layers 200 $\mu$m from the coated surface to minimize both print-through as well as potential coating ablation (which we did not observe anywhere for these mirrors). M4 is thicker than the others, which could have two effects. First, a stiffer substrate requires more stress to correct, and more stress may create more print-through. Second, focusing near the aluminum coating requires propagating through nearly twice as much glass than for M1-M3. The focal region may be expanded due to spherical aberration, or our focal depth estimate may be inaccurate at such depth. Either of these could push the focus closer to the coated surface than we intended. In any case, this nanometer-scale print-through may be avoided by adjusting the focus depth.

\begin{table}
\centering
\begin{threeparttable}
    \caption{WLI measurements before (after) correction [$\mu$m RMS]}
    \label{tab:WLIResults}
        \begin{tabular}{m{1.5cm}| m{2cm} m{2cm} m{2cm} m{2cm}}
            Position & M1&	M2&	M3&	M4\\
            \hline
            1& 1.09 (1.20) & 1.08 (0.99)& 1.79 (1.52)& 0.63 (0.91)\\
            2& 0.98 (1.02) & 0.57 (0.64)& 1.30 (1.25)& 0.99 (1.19)\\
            3& 1.12 (1.11) & 0.63 (0.62)& 0.98 (0.94)& 1.10 (1.07)\\
            4& 1.35 (1.47) & 0.48 (0.53)& 1.58 (1.34)& 0.49 (0.88)\\
            5& 1.01 (1.15) & 0.61 (0.58)& 1.96 (2.06)& 0.65 (1.18)\\
            \hline
            Mean$^a$& 0.066 & 0.025& 0.003& 0.135\\
            \hline
        \end{tabular}
        \begin{tablenotes}
            \small
            \item $^a$ RMS$_{init}-$RMS$_{corr}$ averaged over all 25 fields.
        \end{tablenotes}
    \end{threeparttable}
\end{table}

In addition to correction accuracy, material removal rate (MRR) is an important parameter for manufacturing optical surfaces. Ultrafast laser stress figuring does not remove material, but it does cause deformation to change the surface shape, with an apparent material removal rate (aMRR). If the Zernike fit of the surface is $w$, then the volume to be removed is
\begin{align}
    V=\int_0^{2\pi}{\int_0^R{w(r,\theta) r dr d\theta}} = a_0^0 \pi R^2,
\end{align}
since the Zernike polynomials are mutually orthonormal over a unit disk. $a_0^0$ is the piston term of the Zernike fit and can be arbitrarily set to $a_0^0=-\min{(w)}$ to be consistent with material removal processes, which remove material to the lowest point on the surface (i.e. most negative surface height by convention). The apparent material removal rate is $aMRR=-\min{(w)} \pi R^2 / t,$ where $t$ is the laser processing time. Table \ref{tab:ProcessThroughput} summarizes the process time, aMRR, and average power. These aMRR values, and the correction accuracy (Table \ref{tab:FizeauResults}), are roughly on par with MRF and CCP\cite{suratwala_materials_2018}. The lower aMRR for M3 is because it is mostly composed of higher-order Zernike components that require large stress to correct. The lower aMRR for M4 is because it is thicker and therefore requires more stress. The average laser power is > 100$\times$ smaller than the power available from our ultrafast laser. We expect ULSF could be scaled to produce much higher aMRR values with similar accuracy by using mature high-speed laser processing technologies such as scan mirrors, spatial-light modulators, or diffractive elements. 

\begin{table}
\centering
    \caption{Process throughput data}
    \label{tab:ProcessThroughput}
        \begin{tabular}{m{2.8cm}| m{0.8cm} m{0.8cm} m{0.8cm} m{0.8cm}}
            & M1&	M2&	M3&	M4\\
            \hline
            Process time [min]&     83&     115&    133&    321\\
            aMRR [mm$^3$/hr]&        48.9&   37.9&   6.1&    3.8\\
            Avg. power [mW]&        118&    94&     60&     52\\
            \hline
        \end{tabular}
\end{table}

\section{Conclusions}
\label{sec:Conclusions}
We presented a scalable, accurate, and stable process to correct figure errors in fused silica mirrors without significantly affecting mid-spatial frequency errors. Our calibration process is simple, requiring one small mirror with the same substrate material as the corrected mirror, 6 laser spot patterns, and 7 surface height measurements. We introduced an efficient procedure to allocate laser spots throughout the volume of the corrected mirror, leaving space for multiple correction passes, based on the calibration data and the integrated stress maps for figure correction. We then demonstrated the accuracy and versatility of this process by correcting four flat fused silica mirrors with two different thicknesses and two different coatings. We corrected up to 6$^{th}$ order Zernike polynomials and achieved 10-20 nm RMS accuracy for all four mirrors in 2-3 passes, with equivalent material removal rates similar to existing processes. In dielectric-coated mirrors, we found that the corrected mirror surfaces were stable over a month, and we did not observe any pattern print-through. In aluminum-coated mirrors, however, we observed changes in power up to 100 nm RMS that may be due to laser processing or coating instability, and we observed a few instances of pattern print-through with a maximum amplitude of 3 nm. Future work will investigate these effects on metal coatings, and develop methods to increase throughput while maintaining or improving accuracy. 

\section{Backmatter}

\begin{backmatter}
\bmsection{Funding}
National Science Foundation (2121713).

\bmsection{Acknowledgments}
The authors thank Y. Yao, M. Schattenburg, R. Heilmann, D. Kim, M. Esparza, and M. Echter for valuable discussions regarding this work.

\bmsection{Disclosures}
The authors declare no conflicts of interest.

\bmsection{Data availability} Data underlying the results presented in this paper are available in Ref.\cite{chalifoux_arnold_laverty_2022}.

\bmsection{Supplemental document}
See Supplement 1 for supporting content. 

\end{backmatter}

%%%%%%%%%%%%%%%%%%%%%%% References %%%%%%%%%%%%%%%%%%%%%%%%%
\bibliography{ULSF2022}

\begin{thebibliography}{10}
\newcommand{\enquote}[1]{``#1''}

\bibitem{schiesser_effect_2019}
E.~M. Schiesser, A.~Bauer, and J.~P. Rolland, \enquote{Effect of freeform
  surfaces on the volume and performance of unobscured three mirror imagers in
  comparison with off-axis rotationally symmetric polynomials,}
  {\protect\JournalTitle{Opt. Express}} \textbf{27}, 21750--21765 (2019).

\bibitem{rolland_freeform_2021}
J.~P. Rolland, M.~A. Davies, T.~J. Suleski, C.~Evans, A.~Bauer, J.~C.
  Lambropoulos, and K.~Falaggis, \enquote{Freeform optics for imaging,}
  {\protect\JournalTitle{Optica}} \textbf{8}, 161--176 (2021).

\bibitem{jones_optimization_1977}
R.~A. Jones, \enquote{Optimization of computer controlled polishing,}
  {\protect\JournalTitle{Appl. Opt.}} \textbf{16}, 218 (1977).

\bibitem{golini_magnetorheological_1999}
D.~Golini, W.~I. Kordonski, P.~Dumas, and S.~J. Hogan,
  \enquote{Magnetorheological finishing ({MRF}) in commercial precision optics
  manufacturing,} in \emph{Proc. {SPIE},}  vol. 3782 (1999), pp. 80--92.

\bibitem{weiser_ion_2009}
M.~Weiser, \enquote{Ion beam figuring for lithography optics,}
  {\protect\JournalTitle{Nucl. Instr. Meth. in Phys. Res. B}} \textbf{267},
  1390--1393 (2009).

\bibitem{schaefer_basics_2018}
D.~Schaefer, \enquote{Basics of ion beam figuring and challenges for real
  optics treatment,} in \emph{Proc. {SPIE},}  vol. 10829 (2018), p. 1082907.

\bibitem{castelli_rapid_2012}
M.~Castelli, R.~Jourdain, P.~Morantz, and P.~Shore, \enquote{Rapid optical
  surface figuring using reactive atom plasma,}
  {\protect\JournalTitle{Precision Engineering}} \textbf{36}, 467--476 (2012).

\bibitem{wang_rifta_2020}
T.~Wang, L.~Huang, H.~Kang, H.~Choi, D.~W. Kim, K.~Tayabaly, and M.~Idir,
  \enquote{{RIFTA}: {A} {Robust} {Iterative} {Fourier} {Transform}-based dwell
  time {Algorithm} for ultra-precision ion beam figuring of synchrotron
  mirrors,} {\protect\JournalTitle{Sci. Rep.}} \textbf{10}, 8135 (2020).

\bibitem{suratwala_materials_2018}
T.~I. Suratwala, \emph{Materials {Science} and {Technology} of {Optical}
  {Fabrication}} (John Wiley \& Sons, Ltd, Hoboken, USA, 2018), 1st ed.

\bibitem{chalifoux_correcting_2018}
B.~D. Chalifoux, R.~K. Heilmann, and M.~L. Schattenburg, \enquote{Correcting
  flat mirrors with surface stress: analytical stress fields,}
  {\protect\JournalTitle{J. Opt. Soc. Am. A}} \textbf{35}, 1705--1716 (2018).

\bibitem{yao_stress_2015}
Y.~Yao, X.~Wang, J.~Cao, and M.~Ulmer, \enquote{Stress manipulated coating for
  fabricating lightweight {X}-ray telescope mirrors,}
  {\protect\JournalTitle{Opt. Express}} \textbf{23}, 28605--28618 (2015).

\bibitem{yao_thermal_2019}
Y.~Yao, B.~D. Chalifoux, R.~K. Heilmann, and M.~L. Schattenburg,
  \enquote{Thermal oxide patterning method for compensating coating stress in
  silicon substrates,} {\protect\JournalTitle{Opt. Express}} \textbf{27},
  1010--1024 (2019).

\bibitem{patterson_ultralightweight_2013}
K.~Patterson and S.~Pellegrino, \enquote{Ultralightweight deformable mirrors,}
  {\protect\JournalTitle{Appl. Opt.}} \textbf{52}, 5327--5341 (2013).

\bibitem{deroo_deterministic_2018}
C.~T. DeRoo, R.~Allured, V.~Cotroneo, E.~N. Hertz, V.~Marquez, P.~B. Reid,
  E.~D. Schwartz, A.~A. Vikhlinin, S.~Trolier-McKinstry, J.~Walker, T.~N.
  Jackson, T.~Liu, and M.~Tendulkar, \enquote{Deterministic figure correction
  of piezoelectrically adjustable slumped glass optics,}
  {\protect\JournalTitle{J. Astron. Telesc. Instrum. Syst.}} \textbf{4}, 019004
  (2018).

\bibitem{kanno_development_2007}
I.~Kanno, T.~Kunisawa, T.~Suzuki, and H.~Kotera, \enquote{Development of
  {Deformable} {Mirror} {Composed} of {Piezoelectric} {Thin} {Films} for
  {Adaptive} {Optics},} {\protect\JournalTitle{IEEE Journal of Selected Topics
  in Quantum Electronics}} \textbf{13}, 155--161 (2007).

\bibitem{rodrigues_modular_2009}
G.~N.~M. Rodrigues, R.~P. Bastaits, S.~Roose, Y.~Stockman, S.~E. Gebhardt,
  A.~J. Schönecker, P.~Villon, and A.~J. Preumont, \enquote{Modular bimorph
  mirrors for adaptive optics,} {\protect\JournalTitle{Opt. Eng.}} \textbf{48},
  034001 (2009).

\bibitem{vdovin_correction_2008}
G.~Vdovin, O.~Soloviev, A.~Samokhin, and M.~Loktev, \enquote{Correction of low
  order aberrations using continuous deformable mirrors,}
  {\protect\JournalTitle{Opt. Express}} \textbf{16}, 2859--2866 (2008).

\bibitem{shen_stresses_1996}
Y.-L. Shen, S.~Suresh, and I.~A. Blech, \enquote{Stresses, curvatures, and
  shape changes arising from patterned lines on silicon wafers,}
  {\protect\JournalTitle{J. Appl. Phys.}} \textbf{80}, 1388--1398 (1996).

\bibitem{yao_stress_2021}
Y.~Yao, B.~Chalifoux, R.~Heilmann, and M.~Schattenburg, \enquote{Stress tensor
  mesostructures for freeform shaping of thin substrates,}
  {\protect\JournalTitle{arXiv:2108.00575 [physics]}}  (2021). ArXiv:
  2108.00575.

\bibitem{zuo_experiments_2021}
H.~E. Zuo, B.~D. Chalifoux, R.~K. Heilmann, and M.~L. Schattenburg,
  \enquote{Experiments and simulations of femtosecond laser micromachined
  features for correcting thin silicon mirror segments of {X}-ray telescopes,}
  in \emph{Proc. {SPIE},}  vol. 11822 (2021), p. 118220V.

\bibitem{chalifoux_using_2017}
B.~Chalifoux, C.~Burch, R.~K. Heilmann, Y.~Yao, H.~E. Zuo, and M.~L.
  Schattenburg, \enquote{Using ion implantation for figure correction in glass
  and silicon mirror substrates for x-ray telescopes,} in \emph{Proc. {SPIE},}
  vol. 10399 (2017), p. 103991D.

\bibitem{beckmann_figure_2020}
C.~M. Beckmann and J.~Ihlemann, \enquote{Figure correction of borosilicate
  glass substrates by nanosecond {UV} excimer laser irradiation,}
  {\protect\JournalTitle{Opt. Express}} \textbf{28}, 18681--18692 (2020).

\bibitem{beckmann_freeform_2022}
C.~M. Beckmann, L.~J. Richter, and J.~Ihlemann, \enquote{Freeform shaping of
  fused silica substrates via viscous deformation induced by a laser patterned,
  stressed film,} {\protect\JournalTitle{Opt. Express}} \textbf{30}, 6726--6737
  (2022).

\bibitem{wang_deformation_2016}
X.~Wang, Y.~Yao, T.~Liu, C.~Liu, M.~P. Ulmer, and J.~Cao, \enquote{Deformation
  of rectangular thin glass plate coated with magnetostrictive material,}
  {\protect\JournalTitle{Smart Mater. Struct.}} \textbf{25}, 085038 (2016).

\bibitem{chalifoux_figure_2021}
B.~D. Chalifoux, I.~J. Arnold, and K.~A. Laverty, \enquote{Figure {Correction}
  of {Glass} {Mirrors} {Using} {Ultrafast} {Lasers} to {Generate}
  {Controllable} {Stress},} in \emph{{OSA} {Optical} {Fabrication} and
  {Testing} {Technical} {Digest},}  (2021), p. OW4B.2.

\bibitem{graitzer_correcting_2011}
E.~Graitzer, G.~Antesberger, G.~Ben-Zvi, A.~Cohen, V.~Dmitriev, and
  S.~Winkelmeier, \enquote{Correcting image placement errors using registration
  control ({RegC}) technology,} in \emph{Proc. {SPIE},}  vol. 7973 (2011), p.
  797312.

\bibitem{seesselberg_optical_2019}
M.~Seesselberg, V.~Dmitriev, J.~Welte, U.~Stern, T.~Cohen, and E.~Graitzer,
  \enquote{Optical system and method for correcting mask defects using the
  system,}  (2019).

\bibitem{bellouard_stress-state_2016}
Y.~Bellouard, A.~Champion, B.~McMillen, S.~Mukherjee, R.~R. Thomson, C.~Pépin,
  P.~Gillet, and Y.~Cheng, \enquote{Stress-state manipulation in fused silica
  via femtosecond laser irradiation,} {\protect\JournalTitle{Optica}}
  \textbf{3}, 1285--1293 (2016).

\bibitem{zuo_ultrafast_2018}
H.~E. Zuo, B.~D. Chalifoux, R.~K. Heilmann, and M.~L. Schattenburg,
  \enquote{Ultrafast laser micro-stressing for correction of thin fused silica
  optics for the {Lynx} {X}-{Ray} {Telescope} {Mission},} in \emph{Proc.
  {SPIE},}  vol. 10699 (2018), p. 1069954.

\bibitem{gattass_femtosecond_2008}
R.~R. Gattass and E.~Mazur, \enquote{Femtosecond laser micromachining in
  transparent materials,} {\protect\JournalTitle{Nat Photon}} \textbf{2},
  219--225 (2008).

\bibitem{davis_writing_1996}
K.~M. Davis, K.~Miura, N.~Sugimoto, and K.~Hirao, \enquote{Writing waveguides
  in glass with a femtosecond laser,} {\protect\JournalTitle{Opt. Lett.}}
  \textbf{21}, 1729--1731 (1996).

\bibitem{glezer_ultrafast-laser_1997}
E.~N. Glezer and E.~Mazur, \enquote{Ultrafast-laser driven micro-explosions in
  transparent materials,} {\protect\JournalTitle{Appl. Phys. Lett.}}
  \textbf{71}, 882--884 (1997).

\bibitem{shimotsuma_self-organized_2003}
Y.~Shimotsuma, P.~G. Kazansky, J.~Qiu, and K.~Hirao, \enquote{Self-{Organized}
  {Nanogratings} in {Glass} {Irradiated} by {Ultrashort} {Light} {Pulses},}
  {\protect\JournalTitle{Phys. Rev. Lett.}} \textbf{91}, 247405 (2003).

\bibitem{schaffer_bulk_2003}
C.~Schaffer, J.~García, and E.~Mazur, \enquote{Bulk heating of transparent
  materials using a high-repetition-rate femtosecond laser,}
  {\protect\JournalTitle{Appl. Phys. A}} \textbf{76}, 351--354 (2003).

\bibitem{champion_direct_2012}
A.~Champion and Y.~Bellouard, \enquote{Direct volume variation measurements in
  fused silica specimens exposed to femtosecond laser,}
  {\protect\JournalTitle{Opt. Mater. Express}} \textbf{2}, 789--798 (2012).

\bibitem{stoney_tension_1909}
G.~G. Stoney, \enquote{The {Tension} of {Metallic} {Films} {Deposited} by
  {Electrolysis},} {\protect\JournalTitle{Proc. Roy. Soc. London A}}
  \textbf{82}, 172--175 (1909).

\bibitem{suresh_thin_2009}
S.~Suresh and L.~B. Freund, \emph{Thin {Film} {Materials}: {Stress}, {Defect}
  {Formation} and {Surface} {Evolution}} (Cambridge University Press, 2009).

\bibitem{huang_anisotropic_2007}
Y.~Huang, D.~Ngo, X.~Feng, and A.~J. Rosakis, \enquote{Anisotropic, non-uniform
  misfit strain in a thin film bonded on a plate substrate,}
  {\protect\JournalTitle{Interaction and Multiscale Mechanics}} \textbf{1},
  123--142 (2007).

\bibitem{chalifoux_arnold_laverty_2022}
B.~D. Chalifoux, I.~Arnold, and K.~Laverty, \enquote{Ultrafast laser stress
  figuring for accurate deformation of thin mirrors: dataset,}
  \url{10.25422/azu.data.19199192} (2022).

\bibitem{noll_zernike_1976}
R.~J. Noll, \enquote{Zernike polynomials and atmospheric turbulence,}
  {\protect\JournalTitle{J. Opt. Soc. Am.}} \textbf{66}, 207--211 (1976).

\bibitem{scherer_optical_1996}
K.~Scherer, L.~Nouvelot, P.~Lacan, and R.~Bosmans, \enquote{Optical and
  mechanical characterization of evaporated {SiO}\_2 layers {Long}-term
  evolution,} {\protect\JournalTitle{Appl. Opt.}} \textbf{35}, 5067--5072
  (1996).

\end{thebibliography}

\end{document}